\documentclass[graybox]{svmult}

\usepackage{mathptmx}       
\usepackage{helvet}         
\usepackage{courier}        
\usepackage{type1cm}        
\usepackage{makeidx}         
\usepackage{graphicx}        
\usepackage{multicol}        
\usepackage[bottom]{footmisc}

\usepackage{subfigure}
\usepackage{tikz}  
\usetikzlibrary{fit}   
\usepackage{upgreek}
\usepackage{pgfgantt}

\usepackage{soul}
\usepackage{url}

\usetikzlibrary{matrix,shapes,arrows,positioning,chains}
\tikzstyle{process} = [rectangle, minimum width=3cm, minimum height=1cm, text centered, draw=black]
\tikzstyle{arrow} = [thick,->,>=stealth]
\tikzstyle{io} = [trapezium, trapezium left angle=70, trapezium right angle=110, text centered]

\makeindex             

\begin{document}

\title*{Ultrasound Image Classification using ACGAN with Small Training Dataset}
\author{Sudipan Saha and Nasrullah Sheikh}
\authorrunning{Sudipan Saha and Nasrullah Sheikh} 
\institute{Presented at 2020 Third International Symposium on Signal and Image Processing (ISSIP)\\
Sudipan Saha \at Fondazione Bruno Kessler, Trento, Italy, \email{sudipan.saha@gmail.com}
\and Nasrullah Sheikh \at IBM Research-Almaden, San Jose, CA, USA, \email{nasrullah.sheikh@ibm.com}}
%
%
\maketitle

\abstract*{B-mode ultrasound imaging is a popular medical imaging technique. Like other image processing tasks, deep learning has been
used for analysis of B-mode ultrasound images in the  last few years. However, training deep learning models
requires large labeled datasets, which is often unavailable for ultrasound images. The lack of large
labeled data is a bottleneck for the
use of  deep learning  in ultrasound image analysis. 
To overcome this challenge, in this work we exploit Auxiliary Classifier Generative
Adversarial Network (ACGAN) that combines the benefits of data augmentation and transfer learning in the same framework. We conduct experiment on a dataset of 
breast ultrasound images that shows the effectiveness of the proposed approach.}

\abstract{B-mode ultrasound imaging is a popular medical imaging technique. Like other image processing tasks, deep learning has been
used for analysis of B-mode ultrasound images in the  last few years. However, training deep learning models
requires large labeled datasets, which is often unavailable for ultrasound images. The lack of large
labeled data is a bottleneck for the
use of  deep learning  in ultrasound image analysis. 
To overcome this challenge, in this work we exploit Auxiliary Classifier Generative
Adversarial Network (ACGAN) that combines the benefits of data augmentation and transfer learning in the same framework. We conduct experiment on a dataset of 
breast ultrasound images that shows the effectiveness of the proposed approach.}

\keywords{Ultrasound image, Small training dataset, Deep learning.}

\section{Introduction}
\label{sec:intro}
Ultrasound imaging is a widely used medical imaging modality which is regularly used in clinical
applications and many areas of biomedical research.  Most popular ultrasound imaging mode is the B (brightness) mode  which
is performed by sweeping the transmitted ultrasound wave over plane to generate an intensity image.

\par
The progress in image analysis in the last few years can be ascribed to the accessibility of large labeled data that triggered the development of 
deep learning algorithms, especially  Convolutional Neural Networks (CNNs). To train a CNN, generally considerable amount of labeled 
data is required. Collection of such large dataset is not difficult for natural (RGB) images.
Huge number of images are uploaded everyday in social media. However, acquiring large dataset is challenging in context of medical
images, especially for ultrasound images. 
\par
In the applications where only small datasets are available, the CNN based models are often adopted in a 
setting known as transfer learning \cite{torrey2010transfer, saha2019semantic}. In this setting, deep learning based models are trained on large dataset for some task where such dataset
is available. Subsequently, the trained model is applied to a different (however related) problem. This strategy has been applied in different fields including medical, 
image processing. Based on this, transfer learning has been applied for B-mode ultrasound image analysis too \cite{meng2017liver}. However, transfer learning from
computer vision domain to ultrasound domain achieve limited success due to the strong difference in natural images and ultrasound images. Another popular technique to mitigate
lack of sufficient training images is data augmentation \cite{tom2018simulating, saha2016unsupervised} mainly based on Generative
Adversarial Network (GAN) \cite{saha2019unsupervisedIgarss}.  
\par
In this work, we propose to combine the data augmentation and training of classifier in a single collaborative step. Towards this, we propose
a framework by using Auxiliary Classifier GAN (ACGAN) \cite{odena2017conditional} that combines GAN and a traditional image classifier in the same framework and learns robust feature representations for transfer learning \cite{saha2020unsupervised}. 
An ACGAN consists of a generator $\mathbf{G}$ and a discriminator $\mathbf{D}$, similar to a standard GAN. In a standard GAN $\mathbf{D}$ is tasked to 
discriminate between real and synthetic or fake images generated by $\mathbf{G}$. Differently in ACGAN, $\mathbf{D}$ is also tasked to predict the correct class of images.
This mechanism  gains from the exploitation of additional images generated by $\mathbf{G}$ and simultaneously learns class-specific features. We postulate that such a solution is 
optimal for ultrasound image classification, since it is able to jointly perform data augmentation and training of a classifier that can be subsequently used in transfer learning setting.
\par
We briefly discuss the state-of-the-art relevant to this work in Section \ref{sectionRelatedWork}. We detail the proposed method in Section \ref{sectionProposedMethod}. Experimental results are presented in Section \ref{sectionResult}. The work is concluded in Section \ref{sectionConclusion}.

\section{Related Works}
\label{sectionRelatedWork}

Considering the relevance to the proposed work, we briefly discuss the transfer learning and GAN based methods in ultrasound image analysis and ACGAN applications in different domains.

\subsection{Transfer Learning}
\label{sectionRelatedWorkTransferLearning}
Meng \textit{et. al.}  \cite{meng2017liver} presented a liver fibrosis classification framework using transfer learning. They used a VGGNet pre-trained
on ImageNet dataset for  feature extraction in the liver ultrasound images. Liu \textit{et. al.}  \cite{liu2017classification}  transferred the CNN model learned from
ImageNet to ultrasound image dataset for thyroid nodule classification into benign and malignant. 
Byra \textit{et. al.}  \cite{byra2018transfer} presented a transfer learning based method for 
assessing liver disease in B-mode ultrasound images. Their method uses a network trained using ImageNet dataset. The network is used 
to obtain high-level features from the liver image sequences.

\subsection{GAN}
\label{sectionRelatedWorkGan}
GAN based methods are capable of capturing the distribution of data and generating synthetic data without explicitly modeling the probability density function.
This capability has made GAN popular in computer vision community. Following the trend, we see few applications in ultrasound image analysis \cite{tom2018simulating}.
Tom and Sheet \cite{tom2018simulating} proposed a GAN inspired method for  simulation of ultrasound images, especially targeting  intravascular
ultrasound (IVUS) simulation. Peng  \textit{et. al.}  \cite{peng2019real} proposed a real-time B-mode ultrasound image 
simulator which can simulate 15 frames/second. Fujioka \textit{et. al.} \cite{fujioka2019breast} proposed an approach using GAN  for simulating
breast ultrasound images.  Nevertheless, most of the applications of GAN in ultrasound image analysis have been limited to mere image simulation.

\subsection{ACGAN}
\label{sectionRelatedWorkAcgan}
To the best of our knowledge, ACGAN \cite{odena2017conditional} has not been used before in ultrasound image analysis. However, it
has found applications in different computer vision
and satellite image processing tasks. Roy \textit{et. al.} \cite{roy2018semantic} proposed a semi-supervised image classification
framework by exploiting ACGAN. Saha \textit{et. al.} proposed a method for multi-temporal satellite image processing
by exploiting ACGAN to train robust networks for transfer learning \cite{saha2020unsupervised}. ACGAN has also been applied to medical
image analysis, e.g., X-ray image analysis \cite{waheed2020covidgan}.

While the proposed approach implicitly uses transfer learning, differently from approaches in Section \ref{sectionRelatedWorkTransferLearning}, it does not use
network trained on natural images. The proposes approach extends the methods in Section \ref{sectionRelatedWorkGan} by using ACGAN described in Section
\ref{sectionRelatedWorkAcgan}.

\tikzset{
    between/.style args={#1 and #2}{
         at = ($(#1)!0.5!(#2)$)
    }
}
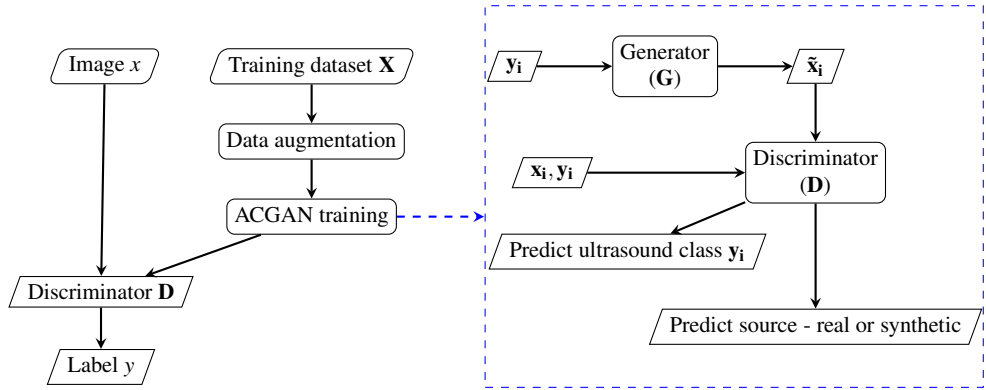
\begin{figure}[!h]
 \centering
 \begin{tikzpicture}
\node (1a) [rounded corners=3pt,draw, io, align=center] {Image $x$};
\node (1b) [rounded corners=3pt,draw, io, align=center, right of=1a,xshift=1.75cm] {Training dataset $\mathbf{X}$};
\node (2b) [rounded corners=3pt,draw,  align=center, below of=1b]{Data augmentation};
\node (3b) [rounded corners=3pt,draw,  align=center, below of=2b]{ACGAN training};
\node (4) [io, draw, align=center, below of=3b, xshift=-2.8cm] {Discriminator $\mathbf{D}$};
\node (5) [io, draw, align=center, below of=4] {Label $y$};

\draw [arrow] (1b) -- (2b);
\draw [arrow] (2b) -- (3b);
\draw [arrow] (3b) -- (4);
\draw [arrow] (1a) -- (4);
\draw [arrow] (4) -- (5);

\node (2RightBlock) [rounded corners=3pt,draw, align=center, right of=1a,xshift=6.45cm] {Generator\\($\mathbf{G}$)};
\node (1RightBlock) [io,draw, left of=2RightBlock, xshift=-1cm] {$\mathbf{y_i}$};
\node (3RightBlock) [io,draw, align=center,right of=2RightBlock, xshift=+1cm] {$\mathbf{\tilde{x}_{i}}$};
\node (4aRightBlock) [rounded corners=3pt,draw, align=center, below of=3RightBlock, yshift=-0.415cm] {Discriminator\\($\mathbf{D}$)};
\node (4bRightBlock) [io,draw, io, align=center, left of=4aRightBlock,xshift=-2.5cm] {$\mathbf{x_{i}},\mathbf{y_i}$};
\node (5aRightBlock) [io,draw, align=center, below of=4aRightBlock, yshift=-1.05cm,xshift=0cm] {Predict source - real or synthetic};
\node (5bRightBlock) [io,draw,  align=center, below of=4aRightBlock, yshift=-0.05cm,xshift=-2.5cm] {Predict ultrasound class $\mathbf{y_i}$};

\draw [arrow] (1RightBlock) -- (2RightBlock);
\draw [arrow] (2RightBlock) -- (3RightBlock);
\draw [arrow] (3RightBlock) -- (4aRightBlock);
\draw [arrow] (4bRightBlock) -- (4aRightBlock);
\draw [arrow] (4aRightBlock) -- (5aRightBlock);
\draw [arrow] (4aRightBlock) -- (5bRightBlock);

\node (ancillaryLeftOf2RightBlock) [left of=2RightBlock, yshift=0.6cm] {};
\node (ancillaryRightOf5aRightBlock) [right of=5aRightBlock, yshift=-0.6cm] {};
\node[draw, blue, dashed,fit= (1RightBlock) (5aRightBlock)(ancillaryLeftOf2RightBlock) (ancillaryRightOf5aRightBlock)]{};

\node (ancillaryRightOf3b) [right of=3b, xshift=1.4cm] {};
\draw [arrow,dashed,blue] (3b) -- (ancillaryRightOf3b);

\end{tikzpicture}
 \caption{Proposed ultrasound image classification framework with small training dataset}
 \label{ultrasoundClassificationFlowchart}
\end{figure}

\section{Proposed Method}
\label{sectionProposedMethod}
Our goal is to classify an ultrasound image $x$ and obtain its corresponding label $y$.  Towards this, we assume that a labeled training dataset (of similar characteristics as $x$)
is available. We denote this dataset as  $\mathbf{X} = \{\mathbf{x_{i}, \forall i=1,...,I}\}$ and their labels as $\mathbf{Y} = \{\mathbf{y_{i}, \forall i=1,...,I}\}$.
However, owing to the lack of labeled ultrasound data, the size $\mathbf{I}$ of the dataset $\mathbf{X}$ is not large enough to train a robust network. To address this problem,
we take advantage of data augmentation. As a first step, we use explicit data augmentation/transformation techniques like rotation, image flipping. Secondly, we implicitly use data augmentation by employing ACGAN. We use dataset $\mathbf{X}$ to train the ACGAN that consists of a discriminator and a generator. After
training, the discriminator of the network is used for classifying $x$. The proposed ultrasound image classification framework is shown in Figure \ref{ultrasoundClassificationFlowchart}.

\subsection{Data Augmentation} 
The training images in $\mathbf{X}$ are processed through popular data augmentation techniques including flipping, rotation (of different angles), noise addition.
The data transformation is applied on the fly with a probability while loading the data for training. Thus, all the training data are not transformed and
only some of the images transformed in each epoch. Moreover, different images are transformed in different epochs. While this serves the purpose of data augmentation,
this also adds a stochastic nature to the training mechanism.  Since data transformation/augmentation is applied on the fly while loading training data, this does not effectively increase
number of training images. Data augmentation is not applied during testing.

\subsection{Training Classifier} We assume that a dataset of $\mathbf{I}$ images is available to train the ACGAN network \cite{odena2017conditional} that 
unifies traditional image classifier and generative adversarial network in same architecture - consisting of the generator $\mathbf{G}$
and the discriminator $\mathbf{D}$ \cite{odena2017conditional}. Images in the training dataset are $\mathbf{X} = \{\mathbf{x_{i}, \forall i=1,...,I}\}$ and its imagewise labels are $\mathbf{Y} = \{\mathbf{y_{i}}$ ($\mathbf{y_{i}} \in \{1,...,\mathbf{C}\}$). Here, number of classes is denoted by $\mathbf{C}$.  After training, discriminator from ACGAN can be used for classification and transfer learning in other ultrasound image analysis. 
\par
Generator $\mathbf{G}$ is tasked to synthesize images indistinguishable from the images in $\mathbf{X}$. Towards this, the generator generates 
class-conditional fake/synthetic images  $\mathbf{\tilde{X}}$. The synthetic images mimic the distribution of $\mathbf{X}$ conditioned
upon class information $\mathbf{\tilde{Y}}$. The generator is implemented using a multi-layered deep network consisting of a series of transposed convolutional layers.
\par
Discriminator $\mathbf{D}$ is tasked to segregate the fake/synthetic images ($\mathbf{\tilde{X}}$) from the ones in $\mathbf{X}$. Additionally, it is 
designed to predict the correct class   $\mathbf{y_{i}}$. Thus, the training mechanism of the discriminator is achieved using a multi-task learning process.
This learning mechanism helps the discriminator to learn useful features to distinguish among different classes and this is implicitly aided
by data augmentation (images generated by $\mathbf{G}$).  The discriminator network is made up of a set of convolutional layers. To further add regularization to the
convolutional layers,  batch normalization layers and leaky ReLU layers are used. The final layer is then fed
to two separate fully connected (FC) layers, corresponding to the prediction of the class and the prediction of the source.  
\par
 Generator ($\mathbf{G}$) is trained in an adversarial fashion in synchronization with discriminator ($\mathbf{D}$). The generator keeps on optimizing its weights
to effectively reproduce the  distribution of target data, while the discriminator keeps on optimizing itself to identify the class of the image ($\mathbf{y_i})$. Additionally,
the discriminator also optimizes its weight to effectively distinguish whether the image is real or synthetic. In this training process, the discriminator has access to the 
images in $\mathbf{X}$, while the generator does not have access to them. The generator updates its weights through its interaction with the discriminator. By
following this mechanism, it can be postulated that both discriminator and generator learns useful semantic features related to the training images. 
\par
The functionality of $\mathbf{D}$ can be decomposed into two functions : $\mathbf{D^{source}}$ (for predicting source as real or fake) and
$\mathbf{D^{class}}$ (to identify the class). To satisfy this, a training mechanism is devised using two objective functions. One of them is  for the likelihood 
related to the source of image ($\mathcal{L}_{source}$). On the other hand, the other objective  is related to the same of the correct class ($\mathcal{L}_{class}$).
While generator tries to maximize $\mathcal{L}_{class}-\mathcal{L}_{source}$, the discriminator tries to  maximize $\mathcal{L}_{class}+\mathcal{L}_{source}$. 

\subsection{Classifying Test Images}
In this step, the weights of discriminator $\mathbf{D}$ are frozen. It is assumed that $\mathbf{D}$ has learned useful semantic features in the training step. Thus, it is
reused for classifying test ultrasound images ($x$).  Between two different functions of $\mathbf{D}$, $\mathbf{D^{source}}$ is unused in this step. The $\mathbf{D^{class}}$ is
used to determine the class of the test images.

\begin{figure}[!t]
\centering

\subfigure{%
            \includegraphics[height=1.6 cm, width=1.6 cm]{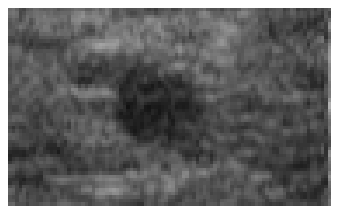}
            \label{imageDataset1}
        }%
\hspace{0.5 cm}
\subfigure{%
            \includegraphics[height=1.6 cm, width=1.6 cm]{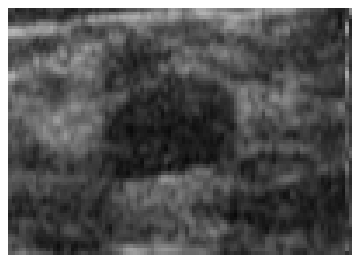}
            \label{imageDataset2}
        }%
\hspace{0.5 cm}
\subfigure{%
            \includegraphics[height=1.6 cm, width=1.6 cm]{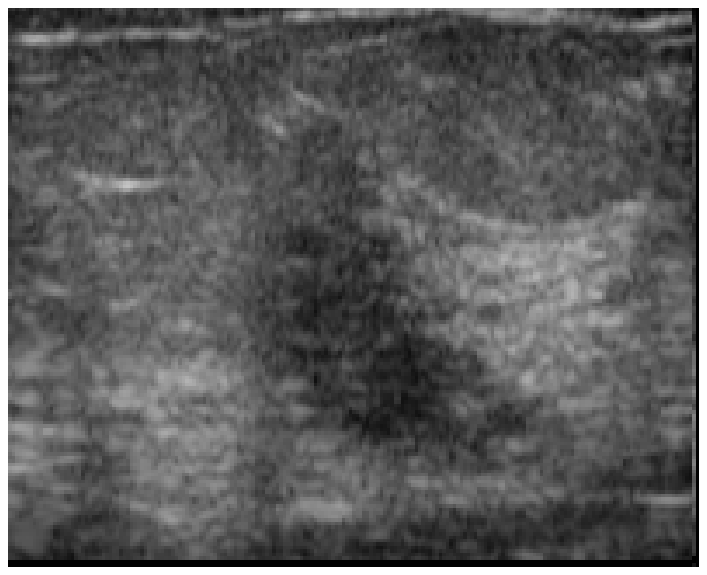}
            \label{imageDataset3}
        }%
\hspace{0.5 cm}
\subfigure{%
            \includegraphics[height=1.6 cm, width=1.6 cm]{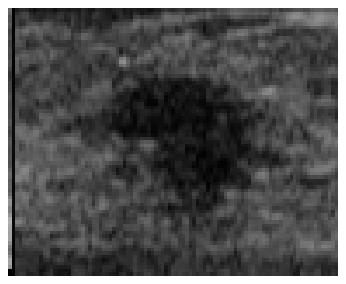}
            \label{imageDataset4}
        }%

\caption{Visualization of some ultrasound images \cite{breastUltrasoundMendelyDataset}.}
\label{figureDatasetImages}
 \end{figure}

\section{Result}
\label{sectionResult}
\subsection{Dataset}
The dataset used for validation is from \cite{breastUltrasoundMendelyDataset}. This dataset consists of
250 ultrasound images from two categories - benign and malignant. 150 images belong to the malignant category, while the
rest belong to benign category. Thus the dataset is considerably small. We show few images  in Figure \ref{figureDatasetImages}. 
The first two images belong to the benign category and the remaining ones belong to the malignant category. It is evident that the two categories do not have 
prominent visual distinction, thus making them challenging to classify.
 
\subsection{Experimental Setup}
For validation purpose, we use 5-fold cross-validation by splitting the dataset of 250 images into 5 splits of 50 image each. For each
evaluation, 4 splits consisting of 200 images are used during the training. For testing, the remaining 50 images are used. In 
this way, 5 independent evaluations are performed and accuracy is averaged over them.
200 training images are much lower than the number of training images generally used in the computer vision tasks.
Thus this dataset is suitable to test the proposed method's capability to deal with lack of sufficient training images in ultrasound image analysis.
\par
We design the discriminator $\mathbf{D}$ using 5 convolutional layers which are fed
to 2 linear layers. Learning rate of 0.0001 is used. 
For optimization, Adam optimizer \cite{kingma2014adam} is used with $\beta_1$ value 0.5. Parameters are chosen
keeping consistency with other works related to ACGAN \cite{odena2017conditional}.  The training procedure of network is performed for 200 epochs.
\par
To experimentally verify whether the model is advantageous in comparison to the existing transfer learning based methods, we designed baseline by using models trained on 
ImageNet \cite{deng2009imagenet}. We used the pre-trained ResNet \cite{he2016deep}, VGGNet \cite{simonyan2014very} models and tuned the final layer. Input images are
resized to the appropriate size before feeding to the pre-trained deep networks. 

\subsection{Quantitative Result}
Quantitative comparison is tabulated in the Table \ref{tableQuantitativeBreastMendeleyClassification}. The accuracy is averaged over the 5-fold validation.
We also show sensitivity (accuracy of detecting positive, i.e., malignant class) and specificity (accuracy of detecting benign class).
The proposed method obtains an accuracy of 98.8\%. Moreover, the proposed method clearly obtains superior result than the transfer learning based methods.
Resnet-18 model obtains an accuracy of 95.6\%. VGG16 and VGG19 models obtain accuracy of 95.6\% and 96.4\%, respectively. Thus the proposed method outperforms
them by atleast 2.4\% or more. Their inferior performance can
be elucidated by the fact that such models were trained for natural images that show different characteristics than ultrasound images. This clearly shows that
learning the classifier using ACGAN learns superior features for ultrasound classification than merely transferring the network.

\renewcommand{\tabcolsep}{2pt}
\begin{table}
\centering
\caption{Quantitative result on the breast dataset from \cite{breastUltrasoundMendelyDataset}}
\begin{tabular}{|c|c|c|c|c|} 
 \hline
\textbf{Method} & Pre-trained & \textbf{Accuracy (\%)} & \textbf{Sensitivity (\%)} & \textbf{Specificity (\%)}  \\ 
\hline
\bf Proposed  & No & 98.80 & 98.67 & 99.00 \\ 
\hline
\bf Resnet-18 & Yes & 95.60 & 96.00 & 95.00 \\ 
\hline
\bf VGG16  & Yes & 95.60 & 95.33 & 96.00 \\ 
\hline
\bf VGG19  & Yes & 96.40 & 96.00 & 97.00 \\ 
\hline
  \end{tabular}
\label{tableQuantitativeBreastMendeleyClassification}
\end{table}

\section{Conclusions}
\label{sectionConclusion}
In this work, we designed an ACGAN based approach for ultrasound image classification. The proposed approach can work with very limited amount of training data by
simultaneous data augmentation by using a generator and training of classifier by using a discriminator. This discriminator is subsequently used for classification of new images, in similar
fashion to transfer learning. However, the proposed approach does not depend on the availability of natural images. Experiment conducted on the breast ultrasound image dataset shows the
superiority of the approach. Moreover, unlike \cite{rodrigues2019cad} the proposed method does not depend on pre-processing steps like speckle reduction and
histogram equalization. The future extension of this work will focus on designing real-time system, incorporating recent deep learning architectures, and consider broader applications.

\bibliographystyle{ieeetr}
\bibliography{ultrasoundClassification}
\end{document}